\documentclass[12pt]{article}
\usepackage{fancyhdr}
\usepackage{url}

 {\fancyhead{}
\fancyfoot[c]{\small{\rule{17.5cm}{1pt}}}}

\usepackage{footnote}
\makesavenoteenv{tabular}
\makesavenoteenv{table}

\usepackage{amsmath}
\usepackage{amssymb}
\usepackage{mathrsfs}
\usepackage{makecell}
\usepackage[]{algorithm2e}
\usepackage{graphicx}
\usepackage{verbatim}
\usepackage{bm}
\usepackage{arydshln}
\usepackage{ dsfont }
\usepackage[square,sort,comma,numbers]{natbib}

\usepackage{color}

\begin{document}
\title{\vspace{-2.5cm}
\begin{center}
\textbf{\small{OPINION}}\\\vspace{-0.5cm} \rule{17.5cm}{1pt}
\end{center}
\vspace{1cm}\textbf{Information Retrieval and Its Sister Disciplines}}

\author{Grace Hui Yang \\
       Department of Computer Science \\
       Georgetown University\\ 
     Washington, D.C., U.S.A. \\ 
       \emph{huiyang@cs.georgetown.edu} \\ 
       \date{}}

\maketitle \thispagestyle{fancy} 

\abstract{ 
   This article presents a summary graph to show the relationships between Information Retrieval (IR) and other related disciplines.  The figure tells the key differences between them and the conditions under which one would transition into another. 
}

\section{Why I drew this graph?}


When I studied Machine Learning (ML), my favorite figure among all was ``The Table of Common Distributions" in \citeauthor*{casella2002statistical}'s 2002 book ``Statistical Inference". It appeared in the book's appendix. Every time when I saw this figure, I was in awe. And I wish someday I could drew one as succinct and expressive as it was. 

Early this year, I had an opportunity to speak about ``Fundamentals of Information Retrieval (IR)" in the first ACM AFIRM Summer School in Cape Town, South Africa.\footnote{\url{http://sigir.org/afirm2019/}} It gave me a chance to reflect what I have learned about IR in these many years of reading, teaching, and research. As time passes by, some of the things, which initially fascinated or puzzled me, no longer seem to be relevant; others last and have repeatedly showed their importance  throughout my interaction with IR. I would like to record, capture and draw them in a figure that I have always wanted to draw. So, at AFIRM 2019, they became one of my presentation slides. 

\section{What does it look like?}
The graph gives a summary of the relationships between IR and its sister disciplines, including Natural Language Processing (NLP), Machine Learning (ML), Question Answering (QA), Artificial Intelligence (AI), Human-Computer Interaction (HCI), Big Data, Databases (DB), Information Science (IS), and Recommendation Systems.  Some of them used to be very close or even part of IR and now have become an independent subject. My guess is when a research subject grows bigger and more sophisticated,  it might be unavoidable for it to split into more specific subjects. And this explains why QA and Recommendation systems appear separated from IR in this graph. Most other nodes, because they have either existed before IR was born, such as AI, HCI, and DB, or have a very established home, such as Distributed systems and Library Science, it is relatively easier to understand why they are separated from IR.

\begin{figure}[t]
    \centering
    \includegraphics[width=0.9\linewidth]{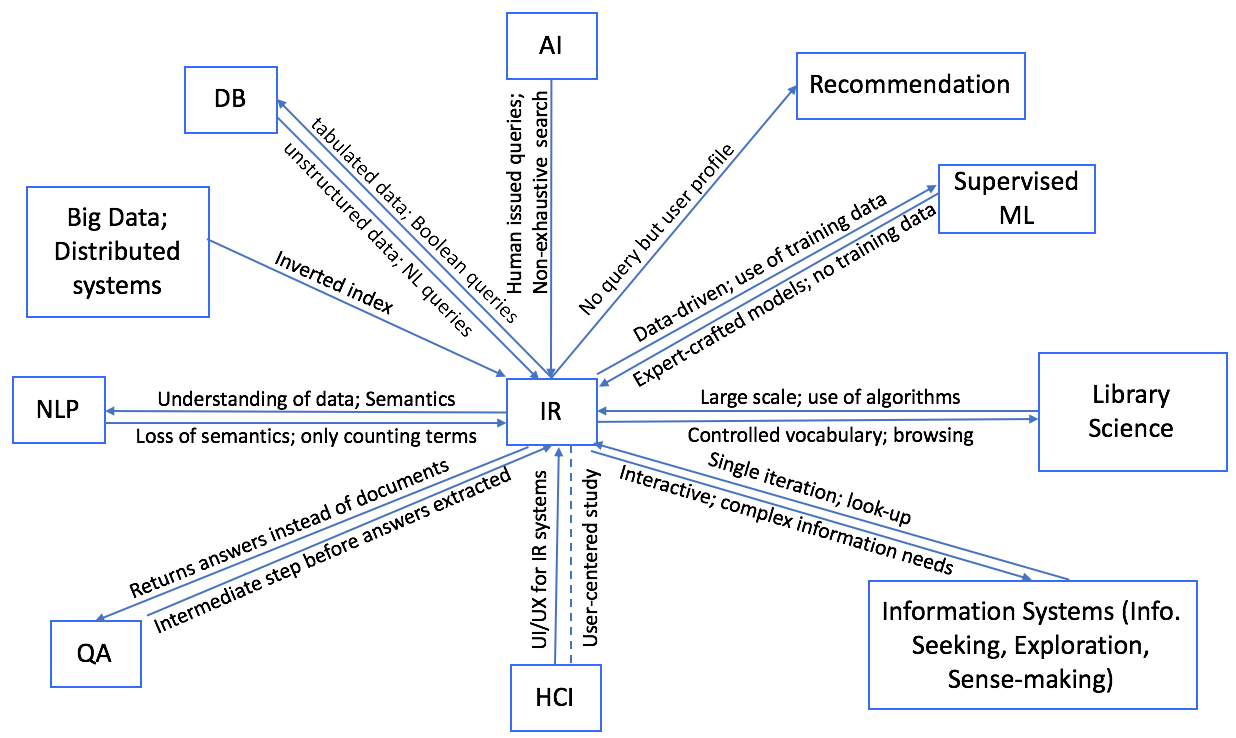}
    \caption{IR and its sister disciplines. Solid lines: transformations or special cases; dashed lines: overlap with.}
    \label{ir}
\end{figure}

The graph should be quite self-explaining. I therefore will only add a few points. 

Many first IR classes probably mention a comparison between DB and IR. It was a good entry point for students in the past since the audience was more familiar with relational databases in their Computer Science undergraduate studies. However, I feel it is a bit difficult to teach Generation Z in this way. It is because they use free text search every day and know more search engines than databases. Nonetheless, I summarize the differences between DB and IR into just two points -- if the data is structured and if the queries are Boolean. 
If we have an imaginary progressive bar to adjust how much structured we would like our data and query to be, then we will find some systems sit in between the two ends. They include semantic Web, faceted search, and other semi-structured searches. 

Among the nodes, the two main economic pillars that generate large revenue for the industry are IR and recommendation. Maybe I should include advertising here, too; but I could not think how different it is from IR and recommendation. In fact, even IR and recommendation are so similar that I can only spot one true difference between them -- whether or not you use a query to search. For other aspects about the two, they are really not that different. Both use large sets of other users' histories to get trained. Both rank. Both personalize their results. Both long for user clicks and engagement. 

Information Science, HCI, and Library Science probably all connect to the human factor and its studies in IR. I feel it is the most difficult and confusing part of the graph. Partly because I am not an expert in those fields, partly because I have worked too much on a tiny sub-field of it -- information seeking. So I think the level of details in this graph might be too much or too few than it should be. I leave it as future work when my understanding in those fields gets improved. 

We are in an exciting moment when there is a lot going on in AI and NLP. For NLP, I think many topics they are interested in, we are interested in, too. The only true and obvious difference, to me, is that NLP puts more emphasis on meanings, semantics, and  understanding of data, and it hopes to gain understanding of text as much as a human can. I think IR cares about those too. Remember we cluster documents into meaningful groups and build search engines on top of them? Remember we add not only the words but also the semantic labels, such as Part-of-Speech tags, semantic roles, and named entity categories, into the index? I think there should be another imaginary progress bar between NLP and IR to include those half NLP, half IR fields. 

For AI, my own belief is that it will be more than deep neural networks. So its rapid development has not yet shown its full potential. We would be clearer about what it can eventually achieve in the next 5 to 15 years. Right now, what I see is that IR always have human users in its system and in most cases, the human user is the driving force of the entire process; while AI could be humanless, such as a smart bulb or a self-driving car. In addition, as interesting as it may sound, IR seems to only care about a few good results showing on the top and neither exhaustive search nor the true optimality is of its concern. 

\section{Concluding Remark}
In the most recent government reports of the world's two biggest economies, they had both listed research fields that they would invest in  tons and people should work on. It looks like IR is not on their lists. Maybe we have been out of the center of people's attention. But my passion to the field won't change and in the graph, I still put IR in the center.

\bibliographystyle{plainnat}
\bibliography{reference}

\end{document}